\documentclass[aps,jcp,preprint,superscriptaddress,letterpaper]{revtex4}
\usepackage{graphicx} % for latex2e
\usepackage{amsmath}
\usepackage{amstext}
\usepackage{amsfonts}
\usepackage{amssymb}
\usepackage{color}

\newcommand{\be}{\begin{equation}}
\newcommand{\ee}{\end{equation}}
\newcommand{\bea}{\begin{eqnarray}}
\newcommand{\eea}{\end{eqnarray}}

\newcommand{\eqa}{\begin{equation}}
\newcommand{\eqz}{\end{equation}}
\newcommand{\eqma}{\begin{eqnarray}}
\newcommand{\eqmz}{\end{eqnarray}}

\usepackage{array}
\newcolumntype{R}[1]{>{\raggedleft  \arraybackslash}p{#1}@{} }
\newcolumntype{C}[1]{>{\centering \arraybackslash}p{#1}@{} }

\newcommand{\pd}[2]{\frac{\partial #1}{\partial #2}}
\newcommand{\cm}{$\text{cm}^{-1}$}
\newcommand{\mx}[1]{\boldsymbol{#1}}

\usepackage{color}

\def\aonems{A$_1$}
\def\ems{E}
\def\fonems{F$_1$}
\def\ftwoms{F$_2$}
\def\aonepg{A$_1^{C_{\text{3v}}}$}
\def\epg{E$^{C_{\text{3v}}}$}

\def\basisA{$\mathcal{B}_A$}
\def\basisB{$\mathcal{B}_B$}
\def\basisC{$\mathcal{B}_C$}
\def\basisT{$\mathcal{B}_T$}

\def\som{Supplementary Material}

\begin{document}

\title{%
Full-dimensional (12D) variational vibrational states of 
CH$_4\cdot$F$^-$: interplay of anharmonicity and tunneling%
\footnote{Dedicated to Professor Attila Cs\'asz\'ar on the occasion of his 60th birthday.}
}

\author{Gustavo Avila}
\email{Gustavo_Avila@telefonica.net}

\author{Edit Matyus}
\email{matyuse@caesar.elte.hu }

\affiliation{Institute of Chemistry,
ELTE, E\"otv\"os Lor\'and University,
P\'azm\'any P\'eter s\'et\'any 1/A,
1117 Budapest, Hungary}

\date{\today}

\begin{abstract}
\noindent %
The complex of a methane molecule and a fluoride anion represents a
12-dimensional (12D), four-well vibrational problem 
with multiple large-amplitude motions,
which has challenged the quantum dynamics community for years.
The present work reports  vibrational band origins and tunneling splittings
obtained in a full-dimensional variational vibrational computation 
using the GENIUSH program and the Smolyak quadrature scheme. 
The converged 12D vibrational band origins and tunneling splittings
confirm complementary aspects of 
the earlier full- and reduced-dimensionality studies:
(1) the tunneling splittings are smaller than $0.02$~\cm;
(2) a single-well treatment is not sufficient 
(except perhaps the zero-point vibration) due to a significant anharmonicity
over the wells;
and thus,  
(3) a full-dimensional treatment appears to be necessary.
The present computations extend to a higher energy range than 
earlier work, show that the tunneling splittings increase
upon vibrational excitation of the complex,  
and indicate non-negligible `heavy-atom' tunneling.
\end{abstract}

\maketitle

\section{Introduction}
\noindent % 
The CH$_4\cdot$F$^-$ complex has been subject of experimental interest over the past decades. 
Its infrared spectrum has been recorded and studied
in the methane's stretching region \cite{WiLoBi02,LoWiWiBiGo04,LoWiWiBiLiNjGo06,Ne08,YaGaKiHoThNe12},
and it has been used as a precursor in anion photoelectron spectroscopy
to probe the transition state region of the $\text{F}+\text{CH}_4$ reaction \cite{ChFECzBo11,WeKiNeMa14}.
This experimental activity motivated the computational 
(ro)vibrational quantum dynamics study of the complex \cite{CzBrBo08,WoPaMa12,FaCsCz13}.
This complex also serves as a good prototype for molecular interactions 
with relatively large monomer distortions and strong binding. 

CH$_4\cdot\text{F}^-$ has turned out to be challenging for the current (ro)vibrational methodologies,
due to its high vibrational dimensionality and multi-well, highly anisotropic potential energy landscape.
The vibrational states from Refs. \cite{CzBrBo08}, \cite{WoPaMa12}, and \cite{FaCsCz13},
using the MULTIMODE \cite{MM2}, the MCTDH \cite{MCTDH1,MCTDH2},
and the GENIUSH \cite{MaCzCs09,FaMaCs11} quantum dynamics program packages, respectively,
show several (tens of) wavenumbers (dis)agreement. In the present work, we aim
to resolve this controversy.

There is currently a single, full-dimensional potential energy surface (PES) available 
for the complex developed by Czak\'o, Braams, and Bowman in 2008 \cite{CzBrBo08}, 
which we will refer to as `CBB08 PES'. 
The CBB08 PES was obtained
by fitting permutationally invariant (up to 6th-order) polynomials
to 6547 (plus 3000) electronic energy points of the interaction (plus
fragment) region computed at the frozen-core CCSD(T)/aug-cc-pVTZ level of the 
\emph{ab~initio} theory. The root-mean-square deviation (rmsd) of the fitting, 
within the energy range below 22\,000~\cm, was reported to be 42~\cm, 
and in practice, the PES describes the intermolecular region well up 
to moderate ion-molecule separations.

The equilibrium structure of the complex has $C_{3\text{v}}$ point-group (PG) symmetry with 
the fluoride binding to one of the apexes of the methane tetrahedron. 
Since the F$^-$ anion can bind to any of the four hydrogens of methane, 
there are four equivalent minima on the PES, which are separated by `surmountable' barriers,
and thus, the molecular symmetry (MS) group is $T_\text{d}$(M). 
The complex is bound by $D_\text{e}=2434$~\cm\ on the CBB08 PES \cite{CzBrBo08}, which corresponds
to $D_0=2316$~\cm\ including the zero-point vibrational energy correction, 
and we have found the lowest barrier connecting the equivalent wells
to be $V=1104$~cm$^{-1}$ \emph{(vide infra)}.
It is interesting to compare these values with the similar parameters of prototypical systems 
of hydrogen bonding and tunneling.
The prototype of strong hydrogen bond, (HF)$_2$
features a dissociation energy of $D_0\approx 1050$~\cm\ \cite{QuSu91}, which 
is less than half of the binding energy of CH$_4\cdot\text{F}^-$.
The prototype for (double-well) tunneling, malonaldehyde has a 
barrier of 1410~\cm\ \cite{MiHaTe14}, which is ca.~30\% higher than the barrier in
CH$_4\cdot$F$^-$, but of course, for estimating the tunneling splitting
one has to consider the effective mass and also the shape (width) of the barrier.

The strong interaction of the methane and the fluoride in CH$_4\cdot\text{F}^-$
is accompanied by a relatively large distortion of the methane fragment.
For the interaction (int) region, Table~3 of Ref.~\cite{CzBrBo08} reports 
the equilibrium structure with an elongated C--H bond, 
$r^\text{int}_\text{eq}(\text{C--H}_\text{b})=1.112$~\AA, for the H which binds (b) to the F$^-$,
while for the other three hydrogens, $r^\text{int}_\text{eq}(\text{C--H})=1.095$~\AA.
The corresponding distorted tetrahedral structure 
is characterized by the $\alpha($H--C--H$_\text{b})=110.46^\text{o}$ angle. 
In the CH$_4+$F$^-$ channel, the practically isolated (isol) methane molecule is a regular tetrahedron with 
a C--H equilibrium distance, $r^\text{isol}_\text{eq}(\text{C--H})=1.090$~\AA.

In the forthcoming sections, 
we briefly summarize the quantum dynamics methodology used in this work (Sec.~II), 
explain the symmetry analysis and assignment of the vibrational states (Sec.~III),
and report the vibrational energies obtained in 
the full- and reduced-dimensionality treatments (Sec.~IV). 
After assessment of the convergence of the vibrational energies, 
a detailed comparison is provided with the vibrational energies reported 
in earlier studies \cite{CzBrBo08,WoPaMa12,FaCsCz13}.
The article is concluded (Sec.~V) with the computation of tunneling splittings 
for excited vibrations of the complex up to and slightly beyond the energy range
of the barrier separating the equivalent wells.

\clearpage
\section{Theoretical and computational details \label{ch:varvib}}
\noindent The present work is among the first applications of the GENIUSH--Smolyak 
algorithm and computer program \cite{AvMa19a}. 
The GENIUSH--Smolyak approach combines
the non-product grid (Smolyak) method of Ref.~\cite{tc-gab1}, 
which has been used for several high-dimensional, semi-rigid molecules \cite{tc-gab2,AvCa11b}
as well as molecules with a single large-amplitude motion \cite{LAUVERGNAT201418,NaLa18},
and 
the numerical kinetic energy operator 
(numerical KEO) approach of Ref.~\cite{MaCzCs09}
implemented in the GENIUSH program~\cite{MaCzCs09,FaMaCs11}, 
which includes by now dozens of vibrational-coordinate definitions for floppy systems 
\cite{MaCzCs09,FaMaCs11,FaCsCz13,FaMaCs14,FaSaCs14,SaCs16,SaCsAlWaMa16,SaCsMa17,FaQuCs17,SiSzCs19}.

Concerning the coordinate definition for the fluoride-methane complex, 
we used the $(R,\cos\theta,\phi)$ spherical polar coordinates 
to describe the relative orientation of the methane fragment and the fluoride ion,
and the nine normal coordinates, $q_1,q_2,\ldots,q_9$, 
of methane (the coordinate definition is provided in the \som) 
to describe its internal vibrations.
In the full-dimensional computations, 
we used the KEO given in Eq.~(50) of Ref.~\cite{AvMa19a}, which reads for
the $(\xi_1,\xi_2,\xi_3,\ldots,\xi_D)$ general coordinates with 
the special $\xi_2=c$ choice 
\begin{align}
  \hat{T}^{\rm v}
  &=
  -\frac{1}{2} \sum_{j=1}^{D} 
    \frac{\partial}{\partial c} G_{c,j} \frac{\partial}{\partial \xi_{j}} 
  -\frac{1}{2} \sum_{i=1,i\ne 2}^{D} \sum_{j=1}^{D} 
      G_{i,j} \frac{\partial}{\partial \xi_{i}}\frac{\partial}{\partial \xi_{j}}
  -\frac{1}{2} \sum_{i=1}^{D} B_{i} \frac{\partial}{\partial \xi_{i}}
  + U, 
\label{H12D} \\
B_{i}&= \sum_{k=1,k\ne 2}^{D} \frac{\partial}{\partial \xi_{k}}G_{k,i} \; ,  \nonumber
\end{align}
where $\mx{g}\in{R}^{(D+3)\times(D+3)}$ is the mass-weighted metric tensor, $\mx{G}=\mx{g}^{-1}$,
$\tilde g = \text{det} \mx{g}$, the extrapotential term,
\begin{align}
U
  &= 
  \frac{1}{32}
  \sum_{k=1}^D 
  \sum_{l=1}^D 
  \left[\frac{G_{kl}}{\tilde{g}^2}
  \pd{\tilde{g}}{\xi_k} \pd{\tilde{g}}{\xi_l} + 4 
  \pd{}{\xi_k}\left(\frac{G_{kl}}{\tilde{g}}\pd{\tilde{g}}{\xi_l}\right)\right],\label{eq:extralg}
\end{align}
and the volume element is $\text{d}V=\text{d}\xi_1 \text{d}c \ldots \text{d}\xi_{D}$.
In the full-dimensional treatment of CH$_4\cdot\text{F}^-$,
$D=12$ and the coordinates are $\xi_1=R$, $\xi_2=c=\cos\theta$, $\xi_3=\phi$,
$\xi_{3+i}=q_i$ ($i=1,2,\ldots 9$).
We treat $c=\cos\theta$ differently from the other coordinates 
in Eq.~(\ref{H12D}) in order to avoid a non-symmetric finite basis representation
of the Hamiltonian due to inaccurate integration caused by singular terms in the KEO \cite{AvMa19a}.
The Hamiltonian matrix was constructed using a finite basis representation (FBR) 
for all coordinates except $c$,
for which the sin-cot discrete variable representation (DVR) \cite{SCHIFFEL2010118} 
was used as it is explained in Sec.~IV.E of Ref.~\cite{AvMa19a}. 
The reduced-dimensionality computations have been carried out 
with the original GENIUSH program \cite{MaCzCs09}, using
the Podolsky form of the KEO
(constructed in an automated way for the imposed geometrical constraints) 
and the Hamiltonian matrix was constructed using DVR \cite{LiCa07}.
The lowest eigenvalues and eigenfunctions 
of the Hamiltonian matrix were computed with an iterative Lanczos eigensolver.

Concerning the full-dimensional computations, it is necessary
to reiterate some methodological details from Ref.~\cite{AvMa19a}
and to specify them for the case of the fluoride-methane complex.
First of all, full-dimensional (12D) computations were possible for this complex
because we used normal coordinates for the methane fragment together with 
harmonic oscillator basis functions, which provide a good zeroth-order description.
Hence, the 9D product basis set of the methane fragment can be pruned \cite{WhHa75}, \emph{i.e.,} 
we can discard high-energy basis functions. We used the simple 
\begin{align}
 \sum_{i=1}^9 n_{q_{i}}\le b \; 
 \label{eq:basprun}
\end{align}
pruning condition for the harmonic oscillator indexes, $n_{q_i}$. 
Since several basis functions are discarded from the methane basis set
complying with this condition, 
it is possible to substantially reduce also the number of quadrature points which 
are used to calculate
the overlap and low-order polynomial integrals with the retained basis functions. 
Pruning the grid following this observation was first realized 
by Avila and Carrington \cite{tc-gab1,tc-gab2} in vibrational computations 
using the Smolyak algorithm.

Concerning the 3-dimensional ion-molecule `intermolecular' part, 
described by the $R$, $\cos\theta$, and $\phi$ coordinates, 
we retained the full direct-product basis and grid. 
For $R$, we used a Morse tridiagonal basis set constructed similarly to Ref.~\cite{AvMa19a}, but 
using  the $D_{0}=1975.27$~cm$^{-1}$, $\alpha=0.9$, and 
$\gamma=18$ parameter values which correspond to the 1D cut of the current PES 
(all other coordinates fixed at their equilibrium value).
The $\cos\theta$ degree of freedom was described with sin-cot-DVR basis functions
and quadrature points \cite{SCHIFFEL2010118}, while we used Fourier functions for the 
$\phi$ angle.

We used large basis sets and grids for the intermolecular
degrees of freedom, $(R,\cos\theta,\phi)$, both in the 3D and in the 12D computations,
which is necessary to ensure that the degeneracies (some of them obtained numerically, only) 
and tunneling splittings are well converged \emph{(vide infra).}
In order to converge the energies with respect to the 
intramolecular (methane) part of the basis and grid, 
we have carried out computations with increasing values of 
the $b$ parameter, $b=2,3,$ and 4,
in the basis pruning condition, Eq.~(\ref{eq:basprun}),
and determined a Smolyak grid which integrates exactly the overlap 
and fifth-order polynomials with all basis functions retained 
in the pruned basis. 

Table~\ref{tab:vib} reports the computed vibrational 
energies in comparison with literature values. We estimate the vibrational excitation energies 
from the largest 12D computation, 
in column \basisC, to be converged within 1~\cm\ and the (apparently very small) 
tunneling splittings are converged with an uncertainty lower than 0.05~\cm.
Note that for $b=2$ and $3$ we used only 23 sin-cot-DVR basis functions,
\basisA\ and \basisB, respectively, which results in a 
0.2--0.3~\cm\ (artificial) splitting for some higher excited states, 
but this splitting is reduced to less than $0.05$~\cm\ upon the increase
of the basis set, which is reported in column \basisC\ of the table 
(25~sin-cot-DVR basis functions and $b=4$~basis-pruning parameter).

Finally, we mention that we were able to put together a `fitted' 3D model 
which reproduced the 12D GENIUSH--Smolyak vibrational band origins
with an rmsd of 1.9~\cm\ (`$G_\text{fit}$' in the Table~\ref{tab:vib}).
`$G_\text{fit}$' was obtained by fine-tuning the regular tetrahedral methane structure used 
in the KEO and in the PES.
In this `fitting' procedure, we obtained $r_\text{PES}(\text{C--H})=2.143\, 624$~bohr 
to define the PES cut (which is slightly different from the value used in Ref.~\cite{FaCsCz13}
and reproduces slightly better the 12D stretching excitations of the complex), 
but we had to use a (much!) larger value in the KEO, $r_\text{KEO}(\text{C--H})=2.518\,620$~bohr
(which corresponds to a much smaller effective rotational constant for the methane)
to have a good agreement with the vibrational band origins especially for the bending excitations.

Note that `3D($G_\text{fit}$)' is merely a fitted model, which was designed to reproduce the 12D GENIUSH--Smolyak
vibrational band origins, and for which we deliberately used two adjustable parameters, 
the C--H distance in the KEO and in the PES.
It differs in this two-parameter adjustment scheme from 
the `rigorous' 3D reduced-dimensionality treatment (for which only a single structure is selected)
used in the first 3D computation of CH$_4\cdot$F$^-$ in Ref.~\cite{FaCsCz13}.

\section{Symmetry analysis and assignment of the vibrational states} %~\\
\noindent %
First of all, we assigned the computed states to irreducible representations (irreps) 
of the MS group of the complex. Then, 
we identified single molecular vibrations classified according to the PG symmetry 
of the equilibrium structure as groups of states close in energy (slightly) split due 
the interaction (spread) of the wave function over the equivalent wells.

At the equilibrium structure of $C_{3\text{v}}$ PG symmetry, one of the hydrogens 
of the methane binds to the fluoride anion. Any of the four hydrogen atoms of 
the methane can bind to the fluoride, 
which gives rise to the four equivalent wells and these wells are connected with `surmountable' barriers.
Thereby, the MS group of the complex is $T_\text{d}(\text{M})$, 
for which the symmetry analysis of (the global minimum of) CH$_4\cdot$Ar \cite{FeMa19} 
is almost verbatim adapted.

In order to assign irrep labels to the CH$_4\cdot$F$^-$ vibrational states computed in the present work, 
we analyzed the wave function of the 3D fitted model computations 
(`3D $G_\text{fit}$' column in Table~\ref{tab:vib})
and the labels were transferred to the 12D results based on the energy ordering 
(direct analysis of the 12D hybrid DVR-FBR computation would have been prohibitively expensive). 
We assigned $T_\text{d}(\text{M})$ molecular symmetry labels 
to the 3D wave functions by computing their overlap with 
2D coupled-rotor (CR) functions, labelled with $[j,j]_{00}$ ($j=0,1,\ldots$) \cite{SaCsMa17,FeMa19}, 
where $j$ is the angular momentum quantum number of the methane and the diatom (corresponding
to the relative motion of the center of mass of the methane and the fluoride), 
coupled to a zero total angular momentum state.
The characters and the irrep decomposition of the $\Gamma^\text{CR}(j)$ representation spanned by
the $[j,j]_{00}$ coupled-rotor functions in $T_\text{d}(\text{M})$ are \cite{FeMa19}:
\begin{align}
\Gamma^{\text{CR}}(0) &= \text{A}_1 \; , \nonumber \\
\Gamma^{\text{CR}}(1) &= \text{F}_2 \; , \nonumber \\
\Gamma^{\text{CR}}(2) &= \text{E}   \oplus \text{F}_2 \; , \nonumber \\
\Gamma^{\text{CR}}(3) &= \text{A}_1 \oplus \text{F}_1 \oplus \text{F}_2 \; , \nonumber \\
\Gamma^{\text{CR}}(4) &= \text{A}_1 \oplus \text{E}   \oplus \text{F}_1 \oplus \text{F}_2 \; , \nonumber \\  
\Gamma^{\text{CR}}(5) &= \text{E}   \oplus \text{F}_1 \oplus \text{F}_2  \; , \nonumber \\ 
\Gamma^{\text{CR}}(6) &= \text{A}_1 \oplus \text{A}_2 \oplus \text{E} \oplus 
\text{F}_1 \oplus 2\text{F}_2 \;, \ldots
\label{eq:crd}
\end{align}

The (hindered) relative rotation of the molecule and the ion over the four wells gives rise to 
tunneling splittings of the vibrations, which can be classified by $C_{3\text{v}}$ 
point-group labels of the symmetry of the local minima. (If there was no
interaction between the wells, each vibrational state would be  
4-fold degenerate due to this feature.)
The MS group species within the tunneling manifold 
of the vibrational modes classified by the PG symmetries (irreps) are \cite{FeMa19}
\begin{align}
  \Gamma(\text{A}_1^{{C}_{3\text{v}}} )
  &= 
  \text{A}_1 \oplus \text{F}_2 \; , \nonumber \\
  %
  %%%%%%%%%%%%%%%%%%%%%%%%%%%%%%%%%%%%%%%%%%%%%%%
  %
  \Gamma({\text{A}_2^{{C}_{3\text{v}}} } )
  &= 
  \text{A}_2 \oplus \text{F}_1 \; , \nonumber \\
  %
  %%%%%%%%%%%%%%%%%%%%%%%%%%%%%%%%%%%%%%%%%%%%%%%
  %
  \Gamma( {\text{E}}^{{C}_{3\text{v}}} )
  &= 
  \text{E} \oplus \text{F}_1 \oplus \text{F}_2 \; . 
\label{eq:pgmsg}
\end{align}
Note that we use 
the $C_\text{3v}$ superscript for the PG irreps, \emph{i.e.,}
$\text{A}_1^{C_\text{3v}}$, $\text{A}_2^{C_\text{3v}}$, and $\text{E}^{C_\text{3v}}$,  
in order to distinguish them from the MS group irreps, which are labelled with
$\text{A}_1$, $\text{A}_2$, $\text{E}$, $\text{F}_1$, and $\text{F}_2$.
The result of this analysis for the computed vibrational wave functions 
is summarized in 
the `$\Gamma(\text{MS})$' and `$\Gamma(\text{PG})$' columns of Table~\ref{tab:vib}, respectively.

The fourth column, `$n_R$', of the table gives the index of the wave function
of the 1D model with active $R$ (all other coordinates fixed at their equilibrium value)
for which the 3D wave function has the largest overlap. Hence, `$n_R$' is 
an index for the excitation along the ion-molecule separation, which we were able to unambiguously
assign for all states listed in the table 
(due to the weak coupling of the the radial and angular degrees of freedom in this complex).

We also note that due to the small tunneling splittings, identification of
the PG vibrations as a set of states of similar character and close in energy 
comprising the appropriate MS group species, Eq.~(\ref{eq:pgmsg}), was possible 
without ambiguities for all states listed in the table.
(This may be contrasted with the floppy CH$_4\cdot$Ar complex, for which unambiguous assignment of
the PG vibrations beyond the zero-point state
of the global minimum was hardly possible \cite{FeMa19}.)
Once, the complete tunneling manifold was assigned and the PG symmetry was found, 
we attached a qualitative description to the states (listed in column `Label' in Table~\ref{tab:vib})
based on the nodal structure along the ion-molecule
separation coordinate $R$, \emph{i.e.,} $n$th `stretching' excitation, 
labelled with $v_\text{s}$, $2v_\text{s}$, etc. 
Excitations different from pure stretching excitation 
were termed `bending' in this qualitative description, and were labelled with $v_\text{b}$, 
$2v_\text{b}$, etc., 
or combinations of stretching and bending, $v_\text{s}+v_\text{b}$, etc. 
Similar qualitative labels had been provided in the earlier
studies \cite{WoPaMa12,FaCsCz13}, which we used to compare with our full-dimensional vibrational 
energies.

\section{Computed vibrational states \label{ch:12d}}
\noindent %
The 12D vibrational energies computed in the present work are reported in 
Table~\ref{tab:vib} (column `12D, GENIUSH--Smolyak'), and the best results
are listed in column `\basisC: (25,4)' of the table. 
The level of convergence and comparison with earlier work is addressed in 
the following paragraphs.

We observe a monotonic decrease of the zero-point vibrational energy (ZPVE) 
upon the increase of the $b$ basis pruning parameter, which indicates that 
the computations are almost variational (the sin-cot-DVR basis and grid is large).
Our best ZPVE value is 9791.6~\cm. It is 5.1~\cm\ larger than the 12D MCTDH result,
which the authors of Ref.~\cite{WoPaMa12} claim to be a variational upper bound to the exact ZPVE. 
The 12D MULTIMODE ZPVE \cite{CzBrBo08} is 3.1~\cm\ (8.2~\cm) larger 
than the lowest value obtained by us (by Ref.~\cite{WoPaMa12} using MCTDH).
The totally symmetric ZPV state (A$_1^{C_\text{3v}}$) is split by 
the relative rotation of the methane and the fluoride to
an \aonems\ and an \ftwoms\ symmetry species, Eq.~(\ref{eq:pgmsg}). 
The \basisC\ 12D computation (numerically) reproduces 
the degeneracy of the \ftwoms\ state within 0.001~\cm\ 
and predicts a tunneling splitting (much) smaller than 0.05~\cm.

At the same time, the MCTDH result 
for the ZPV manifold gives four states with energies differing by 
0.1--0.9~\cm\ (up to 1.2--3.4~\cm, depending on the basis set), which suggests
that the (intermolecular) basis or integration grid used 
in Ref.~\cite{WoPaMa12} was too small.
For higher excited states the (artificial) splittings
increase to 2~\cm\ \cite{WoPaMa12}, whereas
our computations, using a large intermolecular basis set and grid, 
indicate that the tunneling splittings are smaller than 0.05~\cm, and we obtain the 
triple degeneracies converged (numerically) better than 0.001~\cm.

The small tunneling splittings obtained in the present work are in agreement with the 
earlier, 3D computations including the intermolecular, $(R,\cos\theta,\phi)$
coordinates as active vibrational degrees of freedom using the GENIUSH program \cite{FaCsCz13}. 
Although our results confirm the small splittings obtained in the 3D computation of Ref.~\cite{FaCsCz13}, 
we observe larger deviations for the vibrational excitation energies (band origins) 
from the 3D results of Ref.~\cite{FaCsCz13}.
The first 7 vibrational excitations have a 23.5~\cm\ root-mean-square deviation (rmsd) 
from our best 12D energies, 
which suggests that monomer (methane) flexibility effects are important.

A better agreement, with an 11.7~\cm\ rmsd, 
is observed for the first 6 vibrational excitations in comparison with 
the 12D MCTDH result \cite{WoPaMa12}, which accounts for the 
flexibility of the methane fragment and the motion of the fluoride over the four wells, 
but it is affected by incomplete convergence (manifested in the artificial splittings).

The three lowest-energy vibrational fundamentals obtained within a 12D but single-well treatment 
with MULTIMODE \cite{CzBrBo08} has a (surprisingly) large, 20.3~\cm\ rmsd, 
which suggests that in spite of the small tunneling splittings, a multi-well treatment is
necessary for which the normal-coordinate representation is inadequate. It should
be noted that the largest deviation, 32.3~\cm, is observed for the 299.9~\cm\ vibration, and 
this is the state which had the worse convergence properties in the MULTIMODE computation
(Table~5 of Ref.~\cite{CzBrBo08}). 

These observations suggest that although tunneling 
(and the corresponding splittings of the vibrational bands) is 
almost negligible up to 725~\cm\ above ZPVE 
under an energy resolution of 0.05~\cm, 
there is a non-negligible anharmonicity due to the quantum mechanical motion over the multiple wells.
For these reasons, a well converged, full-dimensional variational treatment,
carried out in the present work, appears to be necessary to capture all quantum dynamical features 
of this strongly bound ion-molecule complex.

\begin{table}
\caption{%
 Vibrational states, $\tilde\nu$ in \cm, of CH$_4\cdot$F$^-$ 
 up to 730~\cm\ above the zero-point vibrational energy (ZPVE), 
 corresponding to the full-dimensional PES of Ref.~\cite{CzBrBo08}. 
 Vibrational energies computed in the present work and 
 values taken from literature are shown together for comparison.
 The largest computed splitting for each vibrational manifold is given 
 in parenthesis after the vibrational energy value.
 The most accurate vibrational band origins (from this work) 
 are in column `\basisC: (25,4)'. 
 \label{tab:vib}
}
{%
\scalebox{0.8}{%
\begin{tabular}{@{}cccc c@{}c c c@{}c c@{}c @{\ \ \ } c@{}c c @{\ \ \ \ \ } c@{}c c@{}c @{}}
\cline{1-18}\\[-0.7cm]
\cline{1-18}\\[-0.7cm]
	& & & &
	&\multicolumn{6}{c}{12D}
	&\multicolumn{2}{c}{12D}      
	&\multicolumn{1}{c}{12D}  
	&\multicolumn{2}{c}{3D} 
	&\multicolumn{2}{c}{3D} \\
	& & & &
	&\multicolumn{6}{c}{GENIUSH--Smolyak}
	&\multicolumn{2}{c}{MCTDH}      
	&\multicolumn{1}{c}{MULTIMODE$^\ast$}  
	&\multicolumn{2}{c}{GENIUSH} 
	&\multicolumn{2}{c}{GENIUSH} \\
	& & & &
	&\multicolumn{6}{c}{[this work]}
	&\multicolumn{2}{c}{B3/B4 \cite{WoPaMa12}}      
	&\multicolumn{1}{c}{\cite{CzBrBo08}}  
	&\multicolumn{2}{c}{\cite{FaCsCz13}} 
	&\multicolumn{2}{c}{[this work]} \\
\cline{6-11} \\[-0.75cm]
    & 
    &  
    &  
    &      
    &\multicolumn{2}{c}{$\mathcal{B}_\text{A}$: $(23,2)$$^\text{e,f}$}      
    &\multicolumn{2}{c}{$\mathcal{B}_\text{B}$: $(23,3)$$^\text{e,f}$}      
    &\multicolumn{2}{c}{\textbf{$\mathcal{B}_\text{C}$: }$\boldsymbol{(25,4)}$$^\text{e,f}$}
    &\multicolumn{2}{c}{}        
    &\multicolumn{1}{c}{} 
    &\multicolumn{2}{c}{} 
    &\multicolumn{2}{c}{} \\
    \#    & 
    $\Gamma(\text{MS})$$^\text{a}$  &  
    $\Gamma(\text{PG})$$^\text{b}$  &  
    $n_R$$^\text{c}$  &      
    Label$^\text{d}$ 
    &\multicolumn{2}{c}{$[3.9\cdot 10^5]$$^\text{e}$}      
    &\multicolumn{2}{c}{$[1.6\cdot 10^6]$$^\text{e}$}      
    &\multicolumn{2}{c}{$[5.6\cdot 10^6]$$^\text{e}$}
    &\multicolumn{2}{c}{}        
    &\multicolumn{1}{c}{} 
    &\multicolumn{2}{c}{} 
    &\multicolumn{2}{c}{$G_\text{fit}$$^\text{f,g}$} \\
\cline{1-18}\\[-0.7cm]
0--3	& \aonems$\oplus$\ftwoms             & \aonepg &  0 & ZPVE                       &	9845.6	&	(0.0)	&          9799.1 & (0.0)	&	9791.6	& (0.0)	& 9786.5 & (0.5) & 9794.7 &	461.0	& (0.0)	&	378.8	& (0.0)	\\
4--7	& \aonems$\oplus$\ftwoms             & \aonepg &  1 & $[v_\text{s}]$             &	193.2	&	(0.0)	&	    193.4 & (0.0)	&	193.6	& (0.0)	&  194.4 & (0.1) & 201.1  &	182.5	& (0.0)	&	194.5	& (0.0)	\\
8--15	& \ems$\oplus$\fonems$\oplus$\ftwoms & \epg    &  0 & $[v_\text{b}]$             &	266.3	&	(0.0)	&	    268.3 & (0.0)	&	267.6	& (0.0)	&  271.7 & (1.1) & 299.9  &	284.5	& (0.0)	&	267.7	& (0.0)	\\
16--19	& \aonems$\oplus$\ftwoms             & \aonepg &  2 & $[2v_\text{s}]$            &	378.0	&	(0.0)	&	    378.7 & (0.0)	&	379.2	& (0.0)	&  380.6 & (0.1) & 391    &	355.8	& (0.0)	&	380.3	& (0.0)	\\
20--27	& \ems$\oplus$\fonems$\oplus$\ftwoms & \epg    &  1 & $[v_\text{s}+v_\text{b}]$  &	452.7	&	(0.0)	&	    454.8 & (0.0)	&	454.3	& (0.0)	&  460.2 & (1.3) &	  &	458.8	& (0.0)	&	455.2	& (0.0)	\\
28--31	& \aonems$\oplus$\ftwoms             & \aonepg &  0 & $[2v_\text{b}]$            &	506.7	&	(0.0)	&	    509.8 & (0.0)	&	509.1	& (0.0)	&  528.7 & (0.4) &	  &	533.0	& (0.1) &	509.8	& (0.0)	\\
32--39	& \ems$\oplus$\fonems$\oplus$\ftwoms & \epg    &  0 & $[2v_\text{b}]$            &	523.5	&	(0.2)$^\text{h}$ &  527.0 & (0.2)$^\text{h}$ &	526.0	& (0.0)	&  545.7 & (2.1) &	  &	555.3	& (0.1)	&	525.6	& (0.0)	\\
40--43	& \aonems$\oplus$\ftwoms             & \aonepg &  3 & $[3v_\text{s}]$            &	555.0	&	(0.0)	&	    556.4 & (0.0)	&	557.5	& (0.0)	&        &       &	  &	519.2	& (0.0)	&	556.8	& (0.0)	\\
44--51	& \ems$\oplus$\fonems$\oplus$\ftwoms & \epg    &  2 & $[2v_\text{s}+v_\text{b}]$ &	630.3	&	(0.0)	&	    632.8 & (0.0)	&	632.7	& (0.0)	&	 &	 &	  &		&	&	633.6	& (0.0)	\\
52--55	& \aonems$\oplus$\ftwoms             & \aonepg &  1 & $[v_\text{s}+2v_\text{b}]$ &	685.7	&	(0.0)	&	    689.0 & (0.0)	&	688.5	& (0.0)	&	 &	 &	  &		&	&	690.2	& (0.0)	\\
56--63	& \ems$\oplus$\fonems$\oplus$\ftwoms & \epg    &  1 & $[v_\text{s}+2v_\text{b}]$ &	702.6	&	(0.2)$^\text{h}$ &  706.2 & (0.2)$^\text{h}$ &	705.5	& (0.0)	&	 &	 &	  &		&	&	705.9	& (0.0)	\\
64--67	& \aonems$\oplus$\ftwoms             & \aonepg &  4 & $[4v_\text{s}]$            &	724.2	&	(0.0)	&	    726.6 & (0.0)	&	728.5	& (0.0)	&	 &	 &	  &		&	&	722.9	& (0.0)	\\
\cline{1-18}\\[-0.7cm]
rmsd$^\text{i}$ 
&& & & & 2.4 & & 0.8 & & 0 & & 11.7 & & 20.3  & 23.5 & & 1.9 & \\%[-0.75cm]
\cline{1-18}\\[-0.7cm]
\cline{1-18}\\[-0.6cm]
\end{tabular}
}}
\end{table}
\clearpage
\noindent \emph{Footnotes to Table~\ref{tab:vib}:}
\begin{flushleft}
$^\text{a}$ %
Symmetry assignment (irrep decomposition) in the $T_\text{d}$(M) molecular symmetry group of the complex. \\
$^\text{b}$ %
Symmetry assignment (irrep) within the $C_\text{3v}$ point group of the equilibrium structure. 
\\
$^\text{c}$ %
Dominant overlap of the wave function with the $n_R$th state of 
a 1-dimensional vibrational model along the $R$ degree of freedom (all other coordinates are fixed 
at their equilibrium value). $n_R=0,1,2,\ldots$ labels the states of this 1D model 
in an increasing energy order. \\
$^\text{d}$ %
Qualitative description based on the nodal structure, overlap with lower-dimensional
models, and symmetry assignment. 
These labels are used to compare the vibrational energies
computed in the present work with earlier results \cite{CzBrBo08,WoPaMa12,FaCsCz13}. \\
$^\text{e}$ %
$(N_c,b)$: short label for indicating the basis set size. The $N_c$ value
gives the number of the sin-cot-DVR functions used for $\cos\theta$ 
and $b$ is the basis pruning parameter,
$\sum_{i=1}^9 n_{q_i}\leq b$.
The $R$ and $\phi$ degrees of freedom are described by $N_R=8$ Morse tridiagonal 
and $N_\phi=39$ Fourier basis functions, respectively. 
The overall basis set size, $[N_R N_{c} N_\phi (b+9)! / b! / 9!]$, 
is also shown. \\
$^\text{f}$ %
In the vibrational computations we used atomic masses, 
$m(\text{H})=1.007\,825\,032\,23 $~u,
$m(\text{C})=12$~u, and 
$m(\text{F})=18.998\,403\,162\,73$~u \cite{NIST}. \\
$^\text{g}$ %
Reduced-dimensionality model with active $(R,\cos\theta,\phi)$ degrees of freedom
fitted to reproduce the 12D GENIUSH-Smolyak \basisC\ result.
%s
The methane was treated as a regular tetrahedron and we used
$r_\text{PES}(\text{C--H})=2.143\,624$~bohr in the PES (to reproduce well the stretches)
and $r_\text{KEO}(\text{C--H})=2.518\,620$~bohr in the KEO,
which was necessary to obtain good bending energies. \\
$^\text{h}$ % 
These splittings disappear upon increase of the intermolecular basis set. \\
$^\text{i}$ % 
Root-mean-square deviation of the vibrational excitation energies (ZPVE not included) 
listed in the table (without considering the splittings) 
from the 12D GENIUSH--Smolyak \basisC\ result. \\
$^\ast$ %
Higher energy vibrations, including the 12 fundamental vibrations, are reported in
Ref.~\cite{CzBrBo08}, but they are beyond the (energy) range of the present GENIUSH--Smolyak
computations. \\[2cm]
\end{flushleft}

\section{Excited-state tunneling manifolds \label{ch:tunnel}}
\noindent %
For the ZPV and the lower-energy vibrations,
we obtained tiny tunneling splittings ($<0.05$~\cm, Sec.~\ref{ch:12d}). 
The question arises whether the energy splitting due to tunneling of 
the heavy fragments (heavier than the hydrogen atom)
becomes more significant upon the vibrational excitation of the complex.
The symmetry species in the tunneling manifold are specified in
Eq.~(\ref{eq:pgmsg}), but of course, the symmetry analysis by itself
does not provide any information about the level energies and the size of the splittings.

The height of the barrier, separating the energy wells, can be indicative
for the energy range above which we may expect larger splittings. 
Ref.~\cite{WoPaMa12} estimated the barrier height connecting the equivalent wells 
to be ca.~1270~\cm, and we have pinpointed a lower barrier height, $V_\text{barrier}=1104$~\cm\ 
at $(\theta,\phi)=(90\ ^{\text{o}},45\ ^{\text{o}})$ (see Figure~2 of Ref.~\cite{FaCsCz13}),
$R=5.745$~bohr, and
$(q_1,q_2,\ldots,q_9)=(0,0,0,0,-0.523,0,0,0,-0.524)$. 
It is an interesting question what the `most appropriate' value for the barrier height is
(which is not an experimental observable) 
for comparison with our variational (fully anharmonic, multi-well) vibrational energies.
In the present work, we have not attempted to include quantum nuclear corrections in the height
and will continue using the purely electronic value.
So, we expect the appearance of `sizeable' splittings at around 
ca.~1100~\cm\ measured from the PES minima, which is ca.~700~\cm\ 
above the intermolecular ZPVE (note that the different intermolecular models have somewhat
different ZPVEs).
Furthermore, it is necessary to remember that the methane's bending vibration becomes important
in the energy range near the isolated methane's value at 1311~\cm\ (above the ZPVE), 
which is an obvious limitation for the applicability of the 3D models
and approaching this energy range will require a more efficient 12D treatment.

\begin{figure}
  \includegraphics[scale=1.]{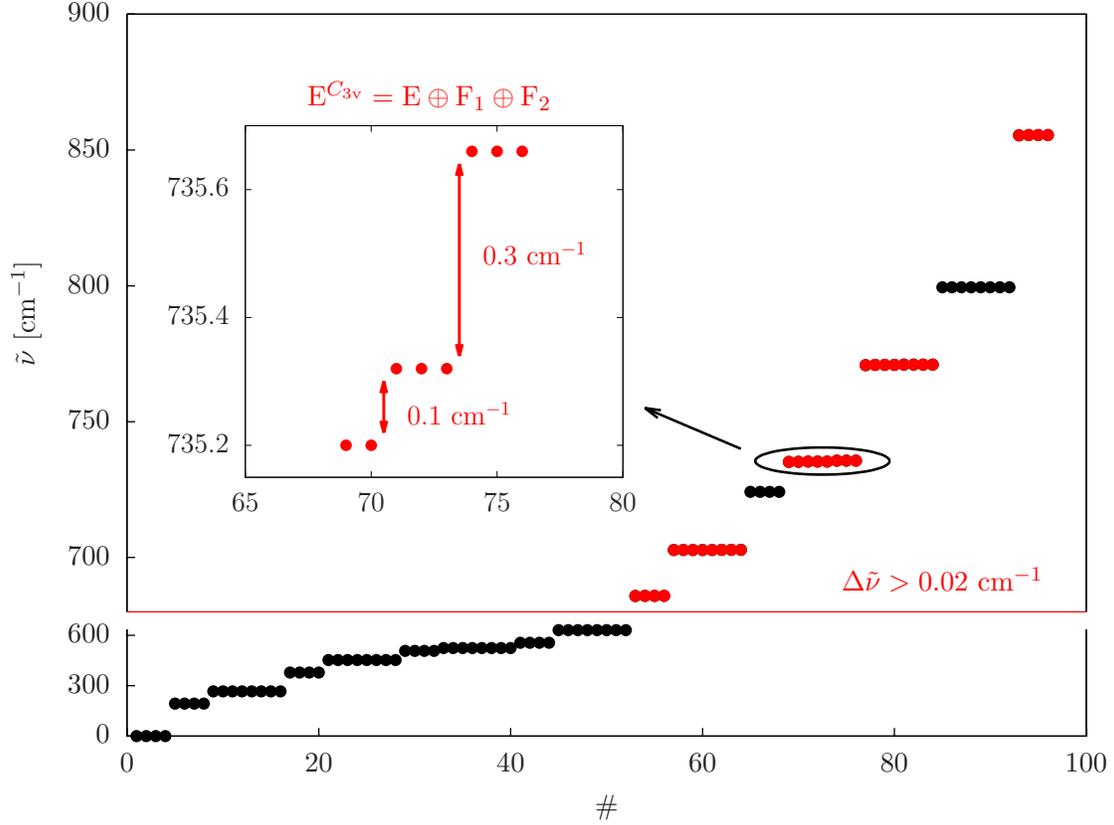}
  \caption{%
    12D vibrational energies measured from the zero-point vibration. 
    Vibrational levels with tunneling splittings, $\Delta\tilde\nu$,  
    larger than 0.02~\cm\ are highlighted in red. 
    The position of the vibrational band origins may have 
    an overall uncertainty as large as 3--4~\cm,
    but the tunneling splittings within a band are converged better than
    0.02~\cm\ (see Figure~\ref{fig:logsplit}).
    \label{fig:split}  
  }
\end{figure}

\begin{figure}
  \includegraphics[scale=1.]{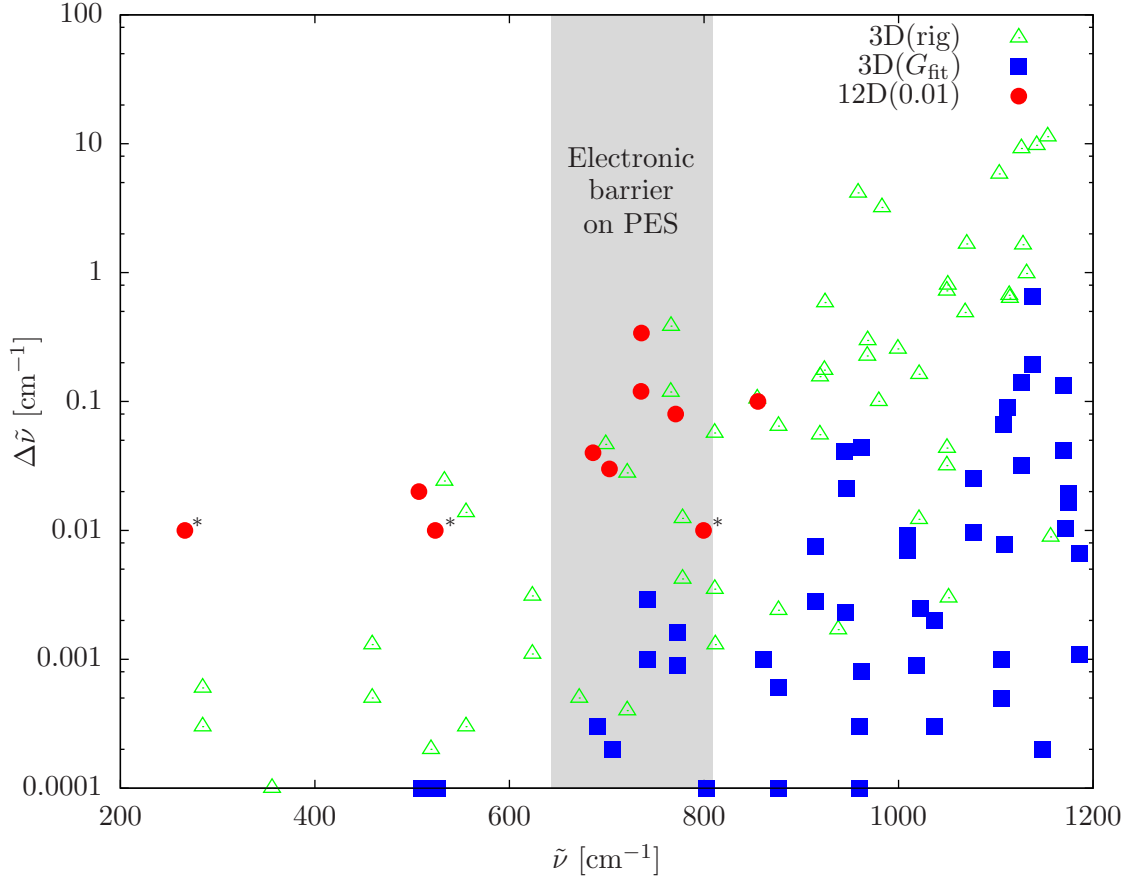}
  \caption{%
  Logarithm of the energy splitting, 
  $\Delta\tilde\nu$ in \cm, among the different symmetry species of the tunneling manifold, 
  Eq.~(\ref{eq:pgmsg}), shown with respect to the $\tilde\nu$ energy of 
  the vibrational state measured from the zero-point vibrational energy (ZPVE).
  The splittings from the rigorous 3D(rig) model, in green triangles, 
  and the fitted 3D($G_\text{fit}$) model, in blue squares, 
  are converged with an uncertainty of 0.000\, 1~\cm.
  The 12D tunneling splittings, in red circles, are converged better than 0.02~\cm\ 
  (the splittings smaller than this value may be affected by numerical artifacts and 
  are labelled with $^\ast$ in the figure).
  The lowest electronic barrier on the CBB08 PES, 1104~\cm\ above the minimum, 
  measured from the intermolecular ZPVE is indicated with the grey shaded area 
  (note that the intermolecular ZPVEs are different for the different models).
  The full list of the computed energies is available in the \som.
  \label{fig:logsplit}  
  }
\end{figure}

In order to explore the tunneling splittings  
and their dependence on the vibrational excitation,
we have carried out extensive 3D and more limited (up to only lower vibrational excitations) 
12D computations with large, (near) saturated angular basis sets \emph{(vide infra).}
The tunneling splittings obtained from the different computations 
are visualized in Figures~\ref{fig:split} and \ref{fig:logsplit}, and 
the energy lists are provided in the \som.

We used the two types of 3D vibrational models
to explore the excited-state tunneling manifold.
First, we have carried out computations with 
the 3D reduced-dimensionality model used already in Ref.~\cite{FaCsCz13}, 
cited in Table~\ref{tab:vib},
which corresponds
to imposing rigorous geometrical constraints in the model through the Lagrangian 
in GENIUSH \cite{MaCzCs09}.
The fixed methane regular tetrahedral structure used in this 3D(rig) model was 
$r(\text{C--H})=1.104$~\AA\ \cite{FaCsCz13}, a value close to the isolated methane's 
$\langle r(\text{C--H}) \rangle_0$ value.
Second, we have carried out computations using the 3D$(G_\text{fit})$ model
designed (`fitted') to minimize the rmsd of the vibrational band origins with respect to 
the 12D GENIUSH--Smolyak results
up to 730~\cm\ (Table~\ref{tab:vib}).
In the largest 3D computations, we used 
31 PO-DVR functions \cite{WeCa92,EcCl92,SzCzNaFuCs02} for $R$ obtained from 
$\mathcal{L}_n^{(\alpha)}$ generalized Laguerre basis functions 
(with $\alpha=2$ and $n=0,1,\ldots,399$) 
and grid points scaled to the $R\in[2,5]$~bohr interval;
61 sin-cot-DVR functions for the $\cos\theta$ degree of freedom,
which was constructed by extending the $\cos(m\theta)$\ ($m=0,...,58$) 
basis set with the $\sin(\theta)$ and the $\sin(2\theta)$ functions \cite{SCHIFFEL2010118,AvMa19a}; 
and 121 Fourier (sine and cosine) functions for the $\phi$ degree of freedom.
This basis is highly saturated and represents a space comparable with
the set of the $Y_j^m(\theta,\phi)$ spherical harmonic functions 
with $j=0,..,60$ and $m=-j,....,j$.
These computations (for both 3D models) reproduce 
the triple degeneracies, which are spatial symmetry features not covered by the DVR grid 
but obtained numerically, within $0.000\ 1$~\cm\ for all computed states.

In the 12D computation of the excited state splittings, we used a 
large angular representation (relative methane-ion rotation) to tightly converge the splittings
but we had to keep the methane basis small (due to the finite computational 
resources), which was sufficient to capture methane's flexibility effects on 
the splittings.
So, we chose $b=2$ in the pruning condition of the methane basis (9D), 
we used 
8~tridiagonal-Morse basis functions and 10~Gauss--Laguerre--Morse points for $R$; 
and a large basis and grid for the angular motion including
31~sin-cot-DVR basis functions and points for $\cos\theta$; and 
45~Fourier basis functions and 48~trapezoidal points for $\phi$.
So, this 12D basis set designed for the computation
of the tunneling splittings (and henceforth labelled with $\mathcal{B}_\text{T}$)
included $(2+9)!/2!/9!\times 8\times 31\times 45=613\,800$ functions
and 
the 12D Smolyak grid ($H=16$, $D=12$ dimensions) 
contained $3\,481$(for the CH$_4$ monomer)$\times 10 \times 31 \times 48 = 51\,797\,280$ 
points to calculate the integrals.
The angular part of \basisT\ is more than 40~\% larger than 
that of \basisC\ in Table~\ref{tab:vib}, which provides 
the current benchmark for the vibrational band origins up to 730~\cm.
Although the uncertainty of the vibrational band origins for \basisT\ is 3--4~\cm\ 
(due to the small $b$ value),
the 12D tunneling splittings are converged with an uncertainty better than 0.02~\cm\ up to 856~\cm,
including 96 vibrational states.

The 12D computations show that 
the tunneling splittings increase
upon the vibrational excitation of the complex
and we find several splittings larger the 0.02~\cm\ uncertainty 
threshold above ca. 680~\cm\ measured from the ZPVE (Figure~\ref{fig:split}).
It is interesting to note that this value is comparable with the value
of the electronic barrier separating the equivalent wells, 
which is 1104~\cm\ from the PES minimum and 
ca. 700~\cm\ measured from the intermolecular ZPVE (different models have different ZPVEs).

Figure~\ref{fig:logsplit} presents the 12D and the 3D splittings.
The 3D$(G_\text{fit})$ model, which reproduces the known 12D vibrational band origins 
very well (Table~\ref{tab:vib}), 
predicts tunneling splittings orders of magnitude smaller than 
the 12D computation (notice the logarithmic scale
in Figure~\ref{fig:logsplit}).
At the same time, the 3D(rig) model (with rigorous geometrical constraints) 
appears to get the splittings right (currently, we have only a few points to compare)
while it has a large error in predicting the vibrational centers (Table~\ref{tab:vib}) 
which may hinder the comparison of the 3D(rig) and 12D splittings at higher energies. 

The differences between the different 3D treatments can be rationalized in terms
of the difference of the `effective' rotational constant of methane in the computation
(the small differences in the PES cut does not affect these results). 
In the 3D treatments, a `smaller' effective methane rotational constant appears to be 
necessary to have good vibrational centers, 3D($G_\text{fit}$),
and a `larger' methane rotational constant is necessary to have good tunneling splittings, 
3D(rig).
Altogether, these observations suggest that in order to have a correct overall description, 
it is necessary to explicitly account for the methane's vibrations together with
the relative rotation of the methane and the fluoride ion.

\clearpage                    
\section{Summary and conclusions}
\noindent Full-dimensional (12D), near-variational, vibrational states are reported for 
the strongly bound complex of the methane molecule and the fluoride anion.
This is the first application of the recently developed 
GENIUSH--Smolyak algorithm and computer program \cite{AvMa19a} 
with a fully coupled, high-dimensional potential energy surface.

Benchmark-quality vibrational band origins are computed with energies 
and tunneling splittings converged better than 1~\cm\ and 0.05~\cm, respectively. 
These computations confirm complementary aspects of earlier studies \cite{CzBrBo08,WoPaMa12,FaCsCz13}, 
which relied on different assumptions or approximations about the dynamics of this complex.
Regarding controversial aspects of earlier work, 
we can confirm that tunneling splittings (up to 730~\cm\ above the zero-point energy)
are small \cite{FaCsCz13}, $<0.05$~\cm, but due to the strong binding and significant monomer distortions, 
a full-dimensional (12D) treatment is necessary \cite{CzBrBo08,WoPaMa12}. 
Although the tunneling splittings are small, 
a single-well (normal coordinate) description \cite{CzBrBo08} is not sufficient due to 
the significant anharmonicity of the multi-well potential energy landscape \cite{WoPaMa12}.
Even so, it is necessary to use a large basis set (and grid) to properly
describe the relative, hindered rotation of the methane fragment and the fluoride anion \cite{FaCsCz13}.

We also show from 12D computations that the tunneling splittings, 
increase with the vibrational excitation of the complex. 
Both the 12D and approximate 3D rigid-monomer computations
indicate that sizeable tunneling splittings, $>0.1$~\cm, 
appear near the `top of the barrier' which separates
the equivalent wells. Interestingly, a `fitted' 3D model,
which gives excellent vibrational band origins, 
predicts orders of magnitude smaller splittings than the 12D computation, while the rigorous, 
reduced-dimensionality 3D model gives reasonable splittings, but 
fails to reproduce the 12D vibrational band energies.

With further progress of the quantum dynamics methodology reported in the present work 
and a potential energy surface with a broader coverage of the nuclear coordinates 
(especially, for the ion-molecule separation),
it will become possible to study 
the effect of the methane vibrational excitation on the tunneling dynamics and 
the intra- to inter-molecular energy transfer under predissociation.

\vspace{1cm}
\noindent\textbf{Supplementary Material} \\ 
Tunneling splittings and vibrational energies obtained from 12D and 3D computations 
and the definition of normal coordinates of methane are provided in the Supplementary Material.
 
\vspace{1cm}
\noindent\textbf{Acknowledgment}\\ 
Financial support of the Swiss National Science Foundation through
a PROMYS Grant (no. IZ11Z0\_166525) is gratefully acknowledged. 
We also thank NIIFI for providing us computer time 
at the Miskolc node of the Hungarian Computing Infrastructure.
EM is thankful to ETH~Z\"urich for supporting a stay as visiting professor during 2019 and 
the Laboratory of Physical Chemistry for their hospitality, where
part of this work has been completed.

% \bibliography{paper.bib} 

\begin{thebibliography}{38}
\expandafter\ifx\csname natexlab\endcsname\relax\def\natexlab#1{#1}\fi
\expandafter\ifx\csname bibnamefont\endcsname\relax
  \def\bibnamefont#1{#1}\fi
\expandafter\ifx\csname bibfnamefont\endcsname\relax
  \def\bibfnamefont#1{#1}\fi
\expandafter\ifx\csname citenamefont\endcsname\relax
  \def\citenamefont#1{#1}\fi
\expandafter\ifx\csname url\endcsname\relax
  \def\url#1{\texttt{#1}}\fi
\expandafter\ifx\csname urlprefix\endcsname\relax\def\urlprefix{URL }\fi
\providecommand{\bibinfo}[2]{#2}
\providecommand{\eprint}[2][]{\url{#2}}

\bibitem[{\citenamefont{Wild et~al.}(2002)\citenamefont{Wild, Loh, and
  Bieske}}]{WiLoBi02}
\bibinfo{author}{\bibfnamefont{D.~A.} \bibnamefont{Wild}},
  \bibinfo{author}{\bibfnamefont{Z.~M.} \bibnamefont{Loh}}, \bibnamefont{and}
  \bibinfo{author}{\bibfnamefont{E.}~\bibnamefont{Bieske}},
  \bibinfo{journal}{Int. J. Mass. Spectrom.} \textbf{\bibinfo{volume}{220}},
  \bibinfo{pages}{273} (\bibinfo{year}{2002}).

\bibitem[{\citenamefont{Loh et~al.}(2004)\citenamefont{Loh, Wilson, Wild,
  Bieske, and Gordon}}]{LoWiWiBiGo04}
\bibinfo{author}{\bibfnamefont{Z.~M.} \bibnamefont{Loh}},
  \bibinfo{author}{\bibfnamefont{R.~L.} \bibnamefont{Wilson}},
  \bibinfo{author}{\bibfnamefont{D.~A.} \bibnamefont{Wild}},
  \bibinfo{author}{\bibfnamefont{E.~J.} \bibnamefont{Bieske}},
  \bibnamefont{and} \bibinfo{author}{\bibfnamefont{M.~S.}
  \bibnamefont{Gordon}}, \bibinfo{journal}{Aust. J. Chem.}
  \textbf{\bibinfo{volume}{57}}, \bibinfo{pages}{1157} (\bibinfo{year}{2004}).

\bibitem[{\citenamefont{Loh et~al.}(2006)\citenamefont{Loh, Wilson, Wild,
  Bieske, Lisy, Njegic, and Gordon}}]{LoWiWiBiLiNjGo06}
\bibinfo{author}{\bibfnamefont{Z.~M.} \bibnamefont{Loh}},
  \bibinfo{author}{\bibfnamefont{L.}~\bibnamefont{Wilson}},
  \bibinfo{author}{\bibfnamefont{D.~A.} \bibnamefont{Wild}},
  \bibinfo{author}{\bibfnamefont{J.}~\bibnamefont{Bieske}},
  \bibinfo{author}{\bibfnamefont{J.~M.} \bibnamefont{Lisy}},
  \bibinfo{author}{\bibfnamefont{B.}~\bibnamefont{Njegic}}, \bibnamefont{and}
  \bibinfo{author}{\bibfnamefont{M.~S.} \bibnamefont{Gordon}},
  \bibinfo{journal}{J. Phys. Chem. A} \textbf{\bibinfo{volume}{110}},
  \bibinfo{pages}{13736} (\bibinfo{year}{2006}).

\bibitem[{\citenamefont{Neumark}(2008)}]{Ne08}
\bibinfo{author}{\bibfnamefont{D.~M.} \bibnamefont{Neumark}},
  \bibinfo{journal}{J. Phys. Chem. A} \textbf{\bibinfo{volume}{112}},
  \bibinfo{pages}{13287} (\bibinfo{year}{2008}).

\bibitem[{\citenamefont{Yacovitch et~al.}(2012)\citenamefont{Yacovitch, Garand,
  Kim, Hock, Theis, and Neumark}}]{YaGaKiHoThNe12}
\bibinfo{author}{\bibfnamefont{T.~I.} \bibnamefont{Yacovitch}},
  \bibinfo{author}{\bibfnamefont{E.}~\bibnamefont{Garand}},
  \bibinfo{author}{\bibfnamefont{J.~B.} \bibnamefont{Kim}},
  \bibinfo{author}{\bibfnamefont{C.}~\bibnamefont{Hock}},
  \bibinfo{author}{\bibfnamefont{T.}~\bibnamefont{Theis}}, \bibnamefont{and}
  \bibinfo{author}{\bibfnamefont{D.~M.} \bibnamefont{Neumark}},
  \bibinfo{journal}{Faraday Discuss.} \textbf{\bibinfo{volume}{157}},
  \bibinfo{pages}{399} (\bibinfo{year}{2012}).

\bibitem[{\citenamefont{Cheng et~al.}(2011)\citenamefont{Cheng, Feng, Du, Zhu,
  Zheng, Czak\'o, and Bowman}}]{ChFECzBo11}
\bibinfo{author}{\bibfnamefont{M.}~\bibnamefont{Cheng}},
  \bibinfo{author}{\bibfnamefont{Y.}~\bibnamefont{Feng}},
  \bibinfo{author}{\bibfnamefont{Y.}~\bibnamefont{Du}},
  \bibinfo{author}{\bibfnamefont{Q.}~\bibnamefont{Zhu}},
  \bibinfo{author}{\bibfnamefont{W.}~\bibnamefont{Zheng}},
  \bibinfo{author}{\bibfnamefont{G.}~\bibnamefont{Czak\'o}}, \bibnamefont{and}
  \bibinfo{author}{\bibfnamefont{J.~M.} \bibnamefont{Bowman}},
  \bibinfo{journal}{J. Chem. Phys.} \textbf{\bibinfo{volume}{134}},
  \bibinfo{pages}{191102} (\bibinfo{year}{2011}).

\bibitem[{\citenamefont{Westermann et~al.}(2014)\citenamefont{Westermann, Kim,
  Weichman, Hock, Yacovitch, Palma, Neumark, and Manthe}}]{WeKiNeMa14}
\bibinfo{author}{\bibfnamefont{T.}~\bibnamefont{Westermann}},
  \bibinfo{author}{\bibfnamefont{J.~B.} \bibnamefont{Kim}},
  \bibinfo{author}{\bibfnamefont{M.~L.} \bibnamefont{Weichman}},
  \bibinfo{author}{\bibfnamefont{C.}~\bibnamefont{Hock}},
  \bibinfo{author}{\bibfnamefont{T.~I.} \bibnamefont{Yacovitch}},
  \bibinfo{author}{\bibfnamefont{J.}~\bibnamefont{Palma}},
  \bibinfo{author}{\bibfnamefont{D.~M.} \bibnamefont{Neumark}},
  \bibnamefont{and} \bibinfo{author}{\bibfnamefont{U.}~\bibnamefont{Manthe}},
  \bibinfo{journal}{Angew. Chem. Int. Ed..} \textbf{\bibinfo{volume}{53}},
  \bibinfo{pages}{1122} (\bibinfo{year}{2014}).

\bibitem[{\citenamefont{Czak\'o et~al.}(2008)\citenamefont{Czak\'o, Braams, and
  Bowman}}]{CzBrBo08}
\bibinfo{author}{\bibfnamefont{G.}~\bibnamefont{Czak\'o}},
  \bibinfo{author}{\bibfnamefont{B.~J.} \bibnamefont{Braams}},
  \bibnamefont{and} \bibinfo{author}{\bibfnamefont{J.~M.}
  \bibnamefont{Bowman}}, \bibinfo{journal}{J. Phys. Chem. A}
  \textbf{\bibinfo{volume}{112}}, \bibinfo{pages}{7466} (\bibinfo{year}{2008}).

\bibitem[{\citenamefont{Wodraszka et~al.}(2012)\citenamefont{Wodraszka, Palma,
  and Manthe}}]{WoPaMa12}
\bibinfo{author}{\bibfnamefont{R.}~\bibnamefont{Wodraszka}},
  \bibinfo{author}{\bibfnamefont{J.}~\bibnamefont{Palma}}, \bibnamefont{and}
  \bibinfo{author}{\bibfnamefont{U.}~\bibnamefont{Manthe}},
  \bibinfo{journal}{J. Phys. Chem. A} \textbf{\bibinfo{volume}{116}},
  \bibinfo{pages}{11249} (\bibinfo{year}{2012}).

\bibitem[{\citenamefont{F\'abri et~al.}(2013)\citenamefont{F\'abri,
  Cs\'asz\'ar, and Czak\'o}}]{FaCsCz13}
\bibinfo{author}{\bibfnamefont{C.}~\bibnamefont{F\'abri}},
  \bibinfo{author}{\bibfnamefont{A.~G.} \bibnamefont{Cs\'asz\'ar}},
  \bibnamefont{and} \bibinfo{author}{\bibfnamefont{G.}~\bibnamefont{Czak\'o}},
  \bibinfo{journal}{J. Phys. Chem. A} \textbf{\bibinfo{volume}{117}},
  \bibinfo{pages}{6975} (\bibinfo{year}{2013}).

\bibitem[{\citenamefont{Bowman et~al.}(2003)\citenamefont{Bowman, Carter, and
  Huang}}]{MM2}
\bibinfo{author}{\bibfnamefont{J.~M.} \bibnamefont{Bowman}},
  \bibinfo{author}{\bibfnamefont{S.}~\bibnamefont{Carter}}, \bibnamefont{and}
  \bibinfo{author}{\bibfnamefont{X.}~\bibnamefont{Huang}},
  \bibinfo{journal}{International Reviews in Physical Chemistry}
  \textbf{\bibinfo{volume}{22}}, \bibinfo{pages}{533} (\bibinfo{year}{2003}).

\bibitem[{\citenamefont{Meyer et~al.}(2009)\citenamefont{Meyer, Gatti, and
  Worth}}]{MCTDH1}
\bibinfo{author}{\bibfnamefont{H.-D.} \bibnamefont{Meyer}},
  \bibinfo{author}{\bibfnamefont{F.}~\bibnamefont{Gatti}}, \bibnamefont{and}
  \bibinfo{author}{\bibfnamefont{G.~A.} \bibnamefont{Worth}},
  \emph{\bibinfo{title}{{MCTDH} for {D}ensity {O}perator}}
  (\bibinfo{publisher}{Wiley-Blackwell}, \bibinfo{year}{2009}),
  chap.~\bibinfo{chapter}{7}, pp. \bibinfo{pages}{57--62}, ISBN
  \bibinfo{isbn}{9783527627400}.

\bibitem[{\citenamefont{Beck et~al.}(2000)\citenamefont{Beck, Jackle, Worth,
  and Meyer}}]{MCTDH2}
\bibinfo{author}{\bibfnamefont{M.}~\bibnamefont{Beck}},
  \bibinfo{author}{\bibfnamefont{A.}~\bibnamefont{Jackle}},
  \bibinfo{author}{\bibfnamefont{G.}~\bibnamefont{Worth}}, \bibnamefont{and}
  \bibinfo{author}{\bibfnamefont{H.-D.} \bibnamefont{Meyer}},
  \bibinfo{journal}{Physics Reports} \textbf{\bibinfo{volume}{324}},
  \bibinfo{pages}{1 } (\bibinfo{year}{2000}), ISSN \bibinfo{issn}{0370-1573}.

\bibitem[{\citenamefont{M\'atyus et~al.}(2009)\citenamefont{M\'atyus, Czak\'o,
  and Cs\'asz\'ar}}]{MaCzCs09}
\bibinfo{author}{\bibfnamefont{E.}~\bibnamefont{M\'atyus}},
  \bibinfo{author}{\bibfnamefont{G.}~\bibnamefont{Czak\'o}}, \bibnamefont{and}
  \bibinfo{author}{\bibfnamefont{A.~G.} \bibnamefont{Cs\'asz\'ar}},
  \bibinfo{journal}{J. Chem. Phys.} \textbf{\bibinfo{volume}{130}},
  \bibinfo{pages}{134112} (\bibinfo{year}{2009}).

\bibitem[{\citenamefont{F\'abri et~al.}(2011)\citenamefont{F\'abri, M\'atyus,
  and Cs\'asz\'ar}}]{FaMaCs11}
\bibinfo{author}{\bibfnamefont{C.}~\bibnamefont{F\'abri}},
  \bibinfo{author}{\bibfnamefont{E.}~\bibnamefont{M\'atyus}}, \bibnamefont{and}
  \bibinfo{author}{\bibfnamefont{A.~G.} \bibnamefont{Cs\'asz\'ar}},
  \bibinfo{journal}{J. Chem. Phys.} \textbf{\bibinfo{volume}{134}},
  \bibinfo{pages}{074105} (\bibinfo{year}{2011}).

\bibitem[{\citenamefont{Quack and Suhm}(1991)}]{QuSu91}
\bibinfo{author}{\bibfnamefont{M.}~\bibnamefont{Quack}} \bibnamefont{and}
  \bibinfo{author}{\bibfnamefont{M.~A.} \bibnamefont{Suhm}},
  \bibinfo{journal}{J. Chem. Phys.} \textbf{\bibinfo{volume}{95}},
  \bibinfo{pages}{28} (\bibinfo{year}{1991}).

\bibitem[{\citenamefont{Mizukami et~al.}(2014)\citenamefont{Mizukami,
  Habershon, and Tew}}]{MiHaTe14}
\bibinfo{author}{\bibfnamefont{W.}~\bibnamefont{Mizukami}},
  \bibinfo{author}{\bibfnamefont{S.}~\bibnamefont{Habershon}},
  \bibnamefont{and} \bibinfo{author}{\bibfnamefont{D.~P.} \bibnamefont{Tew}},
  \bibinfo{journal}{J. Chem. Phys.} \textbf{\bibinfo{volume}{141}},
  \bibinfo{pages}{144310} (\bibinfo{year}{2014}).

\bibitem[{\citenamefont{Avila and M\'atyus}(2019)}]{AvMa19a}
\bibinfo{author}{\bibfnamefont{G.}~\bibnamefont{Avila}} \bibnamefont{and}
  \bibinfo{author}{\bibfnamefont{E.}~\bibnamefont{M\'atyus}},
  \bibinfo{journal}{J. Chem. Phys.} \textbf{\bibinfo{volume}{150}},
  \bibinfo{pages}{174107} (\bibinfo{year}{2019}).

\bibitem[{\citenamefont{Avila and T.~Carrington}(2009)}]{tc-gab1}
\bibinfo{author}{\bibfnamefont{G.}~\bibnamefont{Avila}} \bibnamefont{and}
  \bibinfo{author}{\bibfnamefont{J.}~\bibnamefont{T.~Carrington}},
  \bibinfo{journal}{J. Chem. Phys.} \textbf{\bibinfo{volume}{131}},
  \bibinfo{pages}{174103} (\bibinfo{year}{2009}).

\bibitem[{\citenamefont{Avila and T.~Carrington}(2011{\natexlab{a}})}]{tc-gab2}
\bibinfo{author}{\bibfnamefont{G.}~\bibnamefont{Avila}} \bibnamefont{and}
  \bibinfo{author}{\bibfnamefont{J.}~\bibnamefont{T.~Carrington}},
  \bibinfo{journal}{J. Chem. Phys.} \textbf{\bibinfo{volume}{134}},
  \bibinfo{pages}{054126} (\bibinfo{year}{2011}{\natexlab{a}}).

\bibitem[{\citenamefont{Avila and T.~Carrington}(2011{\natexlab{b}})}]{AvCa11b}
\bibinfo{author}{\bibfnamefont{G.}~\bibnamefont{Avila}} \bibnamefont{and}
  \bibinfo{author}{\bibfnamefont{J.}~\bibnamefont{T.~Carrington}},
  \bibinfo{journal}{J. Chem. Phys.} \textbf{\bibinfo{volume}{134}},
  \bibinfo{pages}{064101} (\bibinfo{year}{2011}{\natexlab{b}}).

\bibitem[{\citenamefont{Lauvergnat and Nauts}(2014)}]{LAUVERGNAT201418}
\bibinfo{author}{\bibfnamefont{D.}~\bibnamefont{Lauvergnat}} \bibnamefont{and}
  \bibinfo{author}{\bibfnamefont{A.}~\bibnamefont{Nauts}},
  \textbf{\bibinfo{volume}{119}}, \bibinfo{pages}{18 } (\bibinfo{year}{2014}),
  ISSN \bibinfo{issn}{1386-1425}.

\bibitem[{\citenamefont{Nauts and Lauvergnat}(2018)}]{NaLa18}
\bibinfo{author}{\bibfnamefont{A.}~\bibnamefont{Nauts}} \bibnamefont{and}
  \bibinfo{author}{\bibfnamefont{D.}~\bibnamefont{Lauvergnat}},
  \bibinfo{journal}{Mol. Phys.} \textbf{\bibinfo{volume}{116}},
  \bibinfo{pages}{3701} (\bibinfo{year}{2018}).

\bibitem[{\citenamefont{F\'abri
  et~al.}(2014{\natexlab{a}})\citenamefont{F\'abri, M\'atyus, and
  Cs\'asz\'ar}}]{FaMaCs14}
\bibinfo{author}{\bibfnamefont{C.}~\bibnamefont{F\'abri}},
  \bibinfo{author}{\bibfnamefont{E.}~\bibnamefont{M\'atyus}}, \bibnamefont{and}
  \bibinfo{author}{\bibfnamefont{A.~G.} \bibnamefont{Cs\'asz\'ar}},
  \bibinfo{journal}{Spectrochim. Acta} \textbf{\bibinfo{volume}{119}},
  \bibinfo{pages}{84} (\bibinfo{year}{2014}{\natexlab{a}}).

\bibitem[{\citenamefont{F\'abri
  et~al.}(2014{\natexlab{b}})\citenamefont{F\'abri, Sarka, and
  Cs\'asz\'ar}}]{FaSaCs14}
\bibinfo{author}{\bibfnamefont{C.}~\bibnamefont{F\'abri}},
  \bibinfo{author}{\bibfnamefont{J.}~\bibnamefont{Sarka}}, \bibnamefont{and}
  \bibinfo{author}{\bibfnamefont{A.~G.} \bibnamefont{Cs\'asz\'ar}},
  \bibinfo{journal}{J. Chem. Phys.} \textbf{\bibinfo{volume}{140}},
  \bibinfo{pages}{051101} (\bibinfo{year}{2014}{\natexlab{b}}).

\bibitem[{\citenamefont{Sarka and Cs\'asz\'ar}(2016)}]{SaCs16}
\bibinfo{author}{\bibfnamefont{J.}~\bibnamefont{Sarka}} \bibnamefont{and}
  \bibinfo{author}{\bibfnamefont{A.~G.} \bibnamefont{Cs\'asz\'ar}},
  \bibinfo{journal}{J. Chem. Phys.} \textbf{\bibinfo{volume}{144}},
  \bibinfo{pages}{154309} (\bibinfo{year}{2016}).

\bibitem[{\citenamefont{Sarka et~al.}(2016)\citenamefont{Sarka, Cs\'asz\'ar,
  Althorpe, Wales, and M\'atyus}}]{SaCsAlWaMa16}
\bibinfo{author}{\bibfnamefont{J.}~\bibnamefont{Sarka}},
  \bibinfo{author}{\bibfnamefont{A.~G.} \bibnamefont{Cs\'asz\'ar}},
  \bibinfo{author}{\bibfnamefont{S.~C.} \bibnamefont{Althorpe}},
  \bibinfo{author}{\bibfnamefont{D.~J.} \bibnamefont{Wales}}, \bibnamefont{and}
  \bibinfo{author}{\bibfnamefont{E.}~\bibnamefont{M\'atyus}},
  \bibinfo{journal}{Phys. Chem. Chem. Phys.} \textbf{\bibinfo{volume}{18}},
  \bibinfo{pages}{22816} (\bibinfo{year}{2016}).

\bibitem[{\citenamefont{Sarka et~al.}(2017)\citenamefont{Sarka, Cs\'asz\'ar,
  and M\'atyus}}]{SaCsMa17}
\bibinfo{author}{\bibfnamefont{J.}~\bibnamefont{Sarka}},
  \bibinfo{author}{\bibfnamefont{A.~G.} \bibnamefont{Cs\'asz\'ar}},
  \bibnamefont{and} \bibinfo{author}{\bibfnamefont{E.}~\bibnamefont{M\'atyus}},
  \bibinfo{journal}{Phys. Chem. Chem. Phys.} \textbf{\bibinfo{volume}{19}},
  \bibinfo{pages}{15335} (\bibinfo{year}{2017}).

\bibitem[{\citenamefont{F\'abri et~al.}(2017)\citenamefont{F\'abri, Quack, and
  Cs\'asz\'ar}}]{FaQuCs17}
\bibinfo{author}{\bibfnamefont{C.}~\bibnamefont{F\'abri}},
  \bibinfo{author}{\bibfnamefont{M.}~\bibnamefont{Quack}}, \bibnamefont{and}
  \bibinfo{author}{\bibfnamefont{A.~G.} \bibnamefont{Cs\'asz\'ar}},
  \bibinfo{journal}{J. Chem. Phys.} \textbf{\bibinfo{volume}{147}},
  \bibinfo{pages}{134101} (\bibinfo{year}{2017}).

\bibitem[{\citenamefont{Simk\'o et~al.}(2019)\citenamefont{Simk\'o,
  Szidarovszky, and Cs\'asz\'ar}}]{SiSzCs19}
\bibinfo{author}{\bibfnamefont{I.}~\bibnamefont{Simk\'o}},
  \bibinfo{author}{\bibfnamefont{T.}~\bibnamefont{Szidarovszky}},
  \bibnamefont{and} \bibinfo{author}{\bibfnamefont{A.~G.}
  \bibnamefont{Cs\'asz\'ar}}, \bibinfo{journal}{J. Chem. Theory Comput.}
  \textbf{\bibinfo{volume}{15}}, \bibinfo{pages}{4156} (\bibinfo{year}{2019}).

\bibitem[{\citenamefont{Schiffel and Manthe}(2010)}]{SCHIFFEL2010118}
\bibinfo{author}{\bibfnamefont{G.}~\bibnamefont{Schiffel}} \bibnamefont{and}
  \bibinfo{author}{\bibfnamefont{U.}~\bibnamefont{Manthe}},
  \bibinfo{journal}{Chem. Phys.} \textbf{\bibinfo{volume}{374}},
  \bibinfo{pages}{118 } (\bibinfo{year}{2010}), ISSN \bibinfo{issn}{0301-0104}.

\bibitem[{\citenamefont{Light and Carrington~Jr.}(2007)}]{LiCa07}
\bibinfo{author}{\bibfnamefont{J.~C.} \bibnamefont{Light}} \bibnamefont{and}
  \bibinfo{author}{\bibfnamefont{T.}~\bibnamefont{Carrington~Jr.}},
  \emph{\bibinfo{title}{Discrete-Variable Representations and their
  Utilization}} (\bibinfo{publisher}{John Wiley \& Sons, Ltd},
  \bibinfo{year}{2007}), pp. \bibinfo{pages}{263--310}.

\bibitem[{\citenamefont{Whitehead and Handy}(1975)}]{WhHa75}
\bibinfo{author}{\bibfnamefont{R.~J.} \bibnamefont{Whitehead}}
  \bibnamefont{and} \bibinfo{author}{\bibfnamefont{N.~C.} \bibnamefont{Handy}},
  \bibinfo{journal}{J. Mol. Spectrosc.} \textbf{\bibinfo{volume}{55}},
  \bibinfo{pages}{356} (\bibinfo{year}{1975}).

\bibitem[{\citenamefont{Ferenc and M\'atyus}(2019)}]{FeMa19}
\bibinfo{author}{\bibfnamefont{D.}~\bibnamefont{Ferenc}} \bibnamefont{and}
  \bibinfo{author}{\bibfnamefont{E.}~\bibnamefont{M\'atyus}},
  \bibinfo{journal}{Mol. Phys.} \textbf{\bibinfo{volume}{117}},
  \bibinfo{pages}{1694} (\bibinfo{year}{2019}).

\bibitem[{\citenamefont{Coursey et~al.}(2015)\citenamefont{Coursey, Schwab,
  Tsai, and Dragoset}}]{NIST}
\bibinfo{author}{\bibfnamefont{J.~S.} \bibnamefont{Coursey}},
  \bibinfo{author}{\bibfnamefont{D.~J.} \bibnamefont{Schwab}},
  \bibinfo{author}{\bibfnamefont{J.~J.} \bibnamefont{Tsai}}, \bibnamefont{and}
  \bibinfo{author}{\bibfnamefont{R.~A.} \bibnamefont{Dragoset}},
  \bibinfo{howpublished}{{Atomic Weights and Isotopic Compositions (version
  4.1): {\tt{http://physics.nist.gov/Comp}} [last accessed on 12 May 2018].
  National Institute of Standards and Technology, Gaithersburg, MD.}}
  (\bibinfo{year}{2015}).

\bibitem[{\citenamefont{Wei and {Carrington, Jr.}}(1992)}]{WeCa92}
\bibinfo{author}{\bibfnamefont{H.}~\bibnamefont{Wei}} \bibnamefont{and}
  \bibinfo{author}{\bibfnamefont{T.}~\bibnamefont{{Carrington, Jr.}}},
  \bibinfo{journal}{J. Chem. Phys.} \textbf{\bibinfo{volume}{97}},
  \bibinfo{pages}{3029} (\bibinfo{year}{1992}).

\bibitem[{\citenamefont{Echave and Clary}(1992)}]{EcCl92}
\bibinfo{author}{\bibfnamefont{J.}~\bibnamefont{Echave}} \bibnamefont{and}
  \bibinfo{author}{\bibfnamefont{D.~C.} \bibnamefont{Clary}},
  \bibinfo{journal}{Chem. Phys. Lett.} \textbf{\bibinfo{volume}{190}},
  \bibinfo{pages}{225} (\bibinfo{year}{1992}).

\bibitem[{\citenamefont{Szalay et~al.}(2003)\citenamefont{Szalay, Czak\'o,
  Nagy, Furtenbacher, and Cs\'asz\'ar}}]{SzCzNaFuCs02}
\bibinfo{author}{\bibfnamefont{V.}~\bibnamefont{Szalay}},
  \bibinfo{author}{\bibfnamefont{G.}~\bibnamefont{Czak\'o}},
  \bibinfo{author}{\bibfnamefont{A.}~\bibnamefont{Nagy}},
  \bibinfo{author}{\bibfnamefont{T.}~\bibnamefont{Furtenbacher}},
  \bibnamefont{and} \bibinfo{author}{\bibfnamefont{A.~G.}
  \bibnamefont{Cs\'asz\'ar}}, \bibinfo{journal}{J. Chem. Phys.}
  \textbf{\bibinfo{volume}{119}}, \bibinfo{pages}{10512}
  (\bibinfo{year}{2003}).

\end{thebibliography}

\end{document}